\documentclass[pdflatex,sn-mathphys-num]{sn-jnl}% Math and Physical Sciences Numbered Reference Style

\usepackage{graphicx}%
\usepackage{multirow}%
\usepackage{amsmath,amssymb,amsfonts}%
\usepackage{amsthm}%
\usepackage{amsmath}%
\usepackage{mathrsfs}%
\usepackage[title]{appendix}%
\usepackage{xcolor}%
\usepackage{textcomp}%
\usepackage{manyfoot}%
\usepackage{booktabs}%
\usepackage{algorithm}%
\usepackage{algorithmicx}%
\usepackage{algpseudocode}%
\usepackage{listings}%
\let\orcidlogo\relax % Clears the existing \orcidlogo definition
\usepackage{orcidlink}%
\theoremstyle{thmstyleone}%
\theoremstyle{thmstyletwo}%

\theoremstyle{thmstylethree}%

\begin{document}

\title[Article Title]{Tetrahedral linkage as an intrinsic measure of glycan antifreeze behavior}


\author*[1,2]{\fnm{Aakash} \sur{Kumar}\email{aakash.kumar@stonybrook.edu}\orcidlink{0000-0003-1338-2530}}%\footnote{Corresponding author: aakash.kumar@stonybrook.edu}
\author[1,2]{\fnm{Shoumik} \sur{Saha}}%\email{iiauthor@gmail.com}

\author*[1,2]{\fnm{Dilip} \sur{Gersappe}}\email{dilip.gersappe@stonybrook.edu}

\affil[1]{\orgdiv{\small{Department of Materials Science and Chemical Engineering}}, \orgname{Stony Brook University}, \orgaddress{\city{Stony Brook}, \postcode{11794}, \state{New York}, \country{USA}}}

\affil[2]{\orgdiv{\small{Institute for Advanced Computational Science}}, \orgname{Stony Brook University},  \orgaddress{\city{Stony Brook}, \postcode{11794}, \state{New York}, \country{USA}}}


\abstract{Antifreeze materials prevent ice-formation by disrupting the ice-formation by binding to certain ice-planes. Cellulose, the most abundant biopolymer, has shown the ability to bind to ice-planes but the exact mechanism of this binding is far from being understood. Molecular dynamics simulations are used to investigate the hydration water of chains of cellulose-type glycans and its significance in the expression of the antifreeze behavior of sugar-derivatives found in some antifreeze materials. We find that glycans are able to prevent water from freezing near its surface by preventing their rearrangement to achieve a highly tetrahedral structure at temperatures well-below the freezing point of water. This validates our hypothesis on the role of tetrahedral coordination based on previous \textit{ab initio} calculations that demonstrated cellulose prefers to bind to ice basal and prismatic planes using a tetrahedral geometry. Our findings suggest that the tetrahedral ordering of water around glycans is the key to understanding and designing cellulose-based antifreeze materials. }




\maketitle

\section*{Impact Statement}\label{impact}

Water is central to all physiological and environmental processes on Earth. In cold regions, its freezing and subsequent thawing have a huge impact on society, as roads and buildings damaged during winter require frequent maintenance and costs. Even more consequential effects are observed in the Arctic and near-Arctic regions, where recurring freeze-thaw cycles destabilize the \textit{permafrost}, imminently threatening the livelihood of its residents. A major goal for the modern built environment in such regions is therefore to make the infrastructure resistant to the ice-water phase change that in recent years has been constantly accelerated due to rising global temperatures. Motivated by our recent findings that cellulose is a promising antifreeze material, we perform molecular dynamics simulations of cellulose-like glycan chains in water to explore its characteristics. This investigation consistently finds that glycans disrupt the tetrahedral arrangement of water during the freezing process, leading to unfrozen supercooled water well below the freezing point of water. Our results therefore lend further support to our previous work, offering cellulose-derived materials as a sustainable antifreeze in the design of infrastructure for a circular economy.

\section{Introduction}\label{sec1}

Antifreeze proteins (AFPs) in organisms that survive extreme cold\cite{balcerzak_designing_2014} have been found to exhibit ice recrystallization inhibition (IRI), where the AFP binds to certain ice planes and impedes the ice-growth. In arctic regions such as Alaska, recurring freeze-thaw cycles prove to be very costly\cite{streletskiy2023costs} as they account for widespread infrastructure damage to buildings and roads, caused by the cyclic expansion/contraction of the water/ice present within the built environment. Synthetic analogs of these AFPs are therefore highly desirable in their infrastructure design as these materials offer a pathway to mitigating such catastrophic structural failures. 

Antifreeze glycoproteins (AFGPs) are another subclass of antifreeze materials with attached glycans where the exact mechanism of their antifreeze activity is far from being understood\cite{deleray2024synthetic}. In our most recent work, we have demonstrated using \textit{ab initio} density functional theory calculations that cellulose, a beta-glycan—as well as the most abundant biopolymer on earth—exhibits potential antifreeze behavior as it can bind to various ice-planes via tetrahedral coordination\cite{kumar_first-principles_2026}. However, design of such materials requires an accurate picture of their interaction with water during the freezing process, which is beyond the reach of first-principles methods. In this letter, we extend our previous work on cellulose-ice interactions to gain an understanding of how glycans affect the water molecules in its vicinity during freezing.  As a natural follow-up to our earlier work\cite{kumar_first-principles_2026} that shed light on how tetrahedral coordination suggested a greater cellulose-ice binding at the \textit{ab initio} level of theory,  in this letter, we demonstrate how the same principle of tetrahedrality in water/ice in the presence of cellulose could be used to characterize the hydration layer of glycans (often referred to as hydration water\cite{bagchi2005water,laage2017water}). To accomplish this task, we utilize classical molecular dynamics (MD) simulations in this work as a necessary step to examine length and time-scales greater than those reached by quantum-mechanical approaches.

\section{Results and Discussion}\label{sec2}

\begin{figure}[ht]
	\centering
	\includegraphics[width=0.5\textwidth]{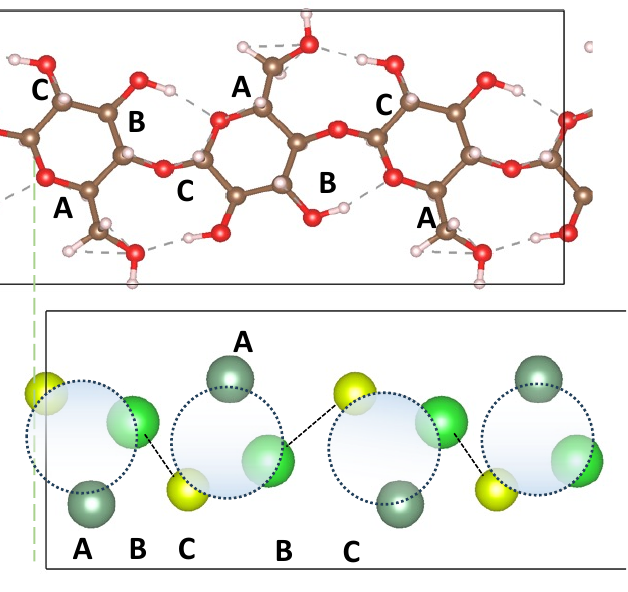}
	\caption{A cg model of the cellulose-type glycan chain used in this work, shown are 4 cg glucose rings, or 2 cellulose repeat units introduced in\cite{kumar_first-principles_2026}}\label{fig1}
\end{figure}

\subsection{Coarse-graining approach}\label{cg}
To model the glycan chains with water, we coarse-grain(cg) each of the atomistic glucose rings using unique A, B, and C beads as shown in fig.\ref{fig1}, which have the same group of atoms to construct the cellulose chain, when repeated. This is in the same spirit as the model for malto-oligosaccharides used by Molinero and Goddard\cite{molinero_m3b_2004}, allowing one to ensure that carbohydrates are not represented as spheroids: carbohydrate mixtures form glasses, while spheres generally form close-packed crystalline solids. With this scheme, we represent the reactive cellulose unit of 42 atoms\cite{kumar_first-principles_2026} used in our recent work using 6 cg (2 A, 2 B, and 2 C) particles. To get statistics for different chain-lengths of glycans in this work, we focus on 3 cases: CG4, 4 equivalent reactive units of cellulose (containing 8 glucose units ); CG6, containing 6 equivalent reactive units of cellulose; and CG8, containing 8 equivalent reactive units of cellulose. 

\subsection{Hydration water}\label{hydration}

It has been widely accepted that the hydration water of AFPs, present near their ice-binding site, is key to understanding their antifreeze behavior\cite{meister2013long}. In the case of a fish AFP found in the antarctic eelpout, it is described as having an ice-like\cite{meister2014observation} arrangement. Nutt and Smith\cite{nutt2008dual} conducted molecular dynamics simulations of an insect AFP found in spruce budworm \textit{Choristoneura fumiferana}, where they established that the hydration water near the ice binding face was more structured with a greater local tetrahedrality than the non-binding planes to facilitate the ice-binding. We use the degree of tetrahedrality ($q$) parameter\cite{chau1998new,errington2001relationship} as in our previous work\cite{kumar_first-principles_2026} to quantify this deviation from the perfect tetrahedral packing of ice. 

The degree of tetrahedrality, \textit{q}, is calculated for each water particle with its 4 nearest-neighbors ($j$,$k$) based on the expression by Errington and Debenedetti\cite{errington2001relationship} shown in eq.\ref{eqn:tetra}
\begin{equation}
	q = 1-\frac{3}{8} \sum_{j=1}^{3} \sum_{k=j+1}^{4}(\cos\psi_{jk}+\frac{1}{3} )^2 \label{eqn:tetra}
\end{equation}

where $\cos\psi_{jk}$ is the angle formed by the water particle at the center with its neighbors ($j$,$k$). Here \textit{q} varies from 0 (disordered/ideal gas) to 1 (perfect tetrahedron). We observe that in all of our simulations, the water near glycan chains is indeed more structured as tetrahedral.

\begin{figure}[ht]
	\centering
	\includegraphics[width=0.9\textwidth]{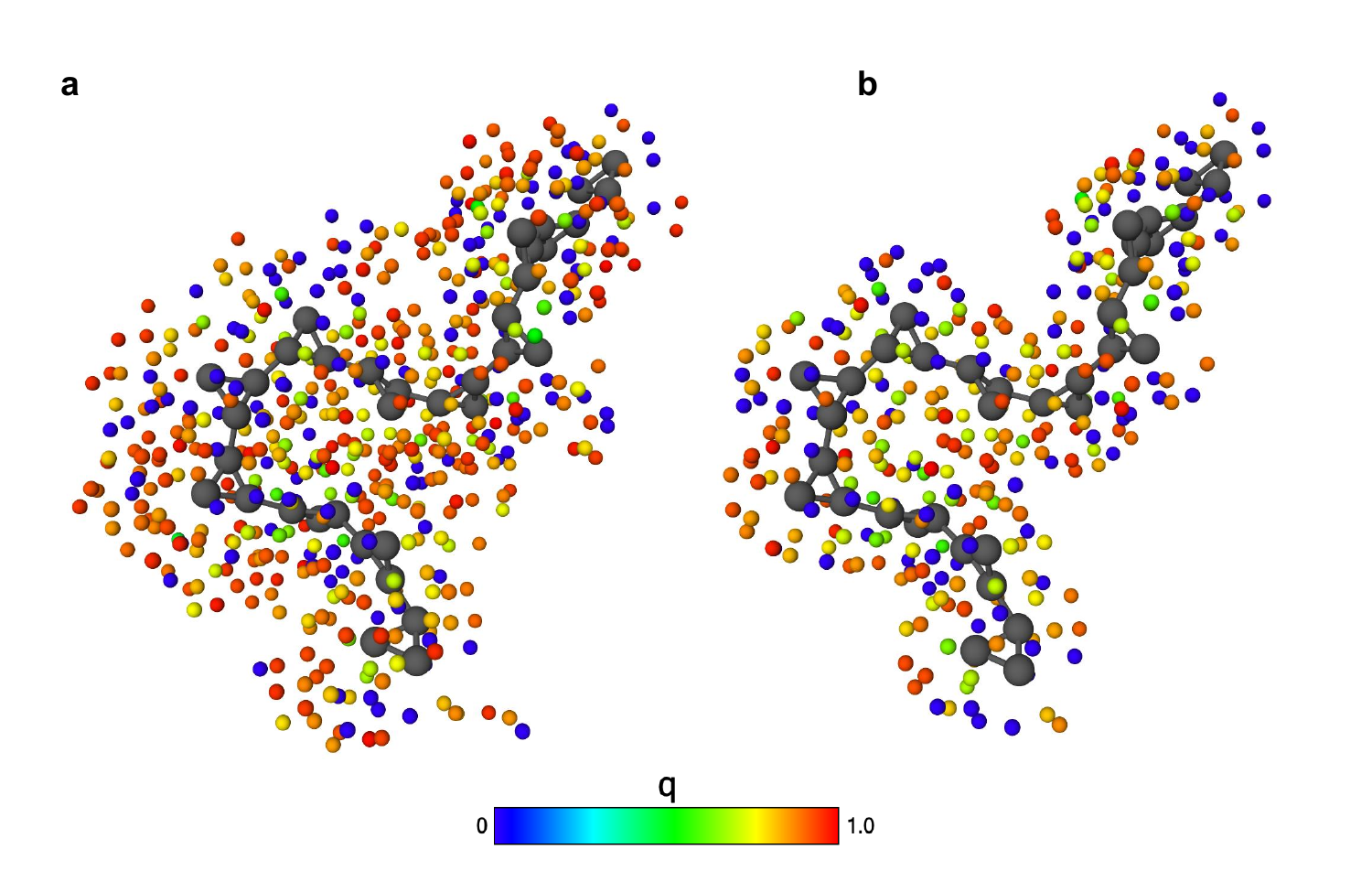}
	\caption{A \textit{q} color-coded snapshot for the water particles surrounding the glycan-chains for a distance of (a)$\le$ 10 \AA~ and (b)  $\le$ 6 \AA~from the chain.}\label{fig2}
\end{figure}

Figure \ref{fig2} elucidates the \textit{q} distribution of the water (after removing the ice phase identified using the chill+ algorithm\cite{nguyen2015identification}) that surrounds the glycan chain with 12 glucose units (CG6). It can be seen in fig.\ref{fig2}a that the water (within a distance of 10~\AA~from the chain) is highly tetrahedral with \textit{q}$\sim$1 suggesting the ``ice-like water'' nature. However, closer to the chain surface ($\le$ 6 \AA), as shown in fig.\ref{fig2}(b), the hydration water presents a greater variation in $q$ suggesting a greater disruption of the water molecules to rearrange to a tetrahedral coordination than seen in fig.\ref{fig2}a. We clarify that several water particles do not have 4 nearest-neighbors within their first hydration shell (3.1~\AA) to be included for the analysis—these are assigned $q$=0, and shown in blue.

%\subsubsection*{Cellulose chain-length}\label{chain_length} 
We compare the effect of increasing the chain-length on the distribution of \textit{q} values for water particles when going from CG4 to CG6 to CG8. Interestingly, we observe that the $q$ distribution for the three cases remains fairly the same suggesting that the effect is highly localized to the glycan-water interface. 

\begin{figure}[h]
	\centering
	\includegraphics[width=1.0\textwidth]{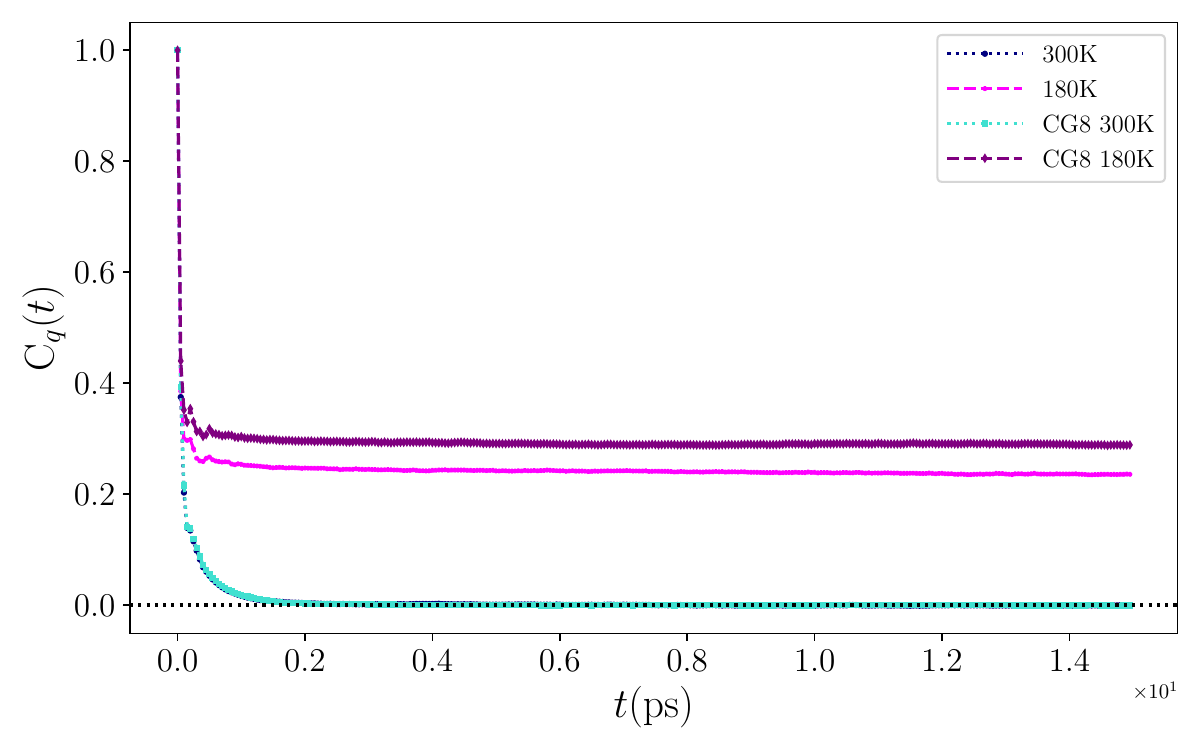}
	\caption{ $C_q(t)$ decay for pure water and the CG8+water systems at 300 K and 180 K}\label{fig3}
\end{figure}

\subsection{\textit{q} autocorrelation}\label{autocorrelation}

To further probe the extent to which the glycan chain modifies the tetrahedrality of water, we compare its time correlation function at 300 K and 180 K as shown in fig.~\ref{fig3}. %It can be observed that  $\left\langle q\right\rangle$ changes roughly from around its peak at 0.8 (liquid water) at 300 K to close to 1 (ice) at 180 K in fig.\ref{fig3} a,b. Figure \ref{fig3} c,d show the distribution of \textit{q} by the number of neighbors (degree) around each water particle. The distribution at 300 K tends to be more random, with the overall \textit{q} close to 0.8. We find that at 180 K, the ice-like tetrahedral character is most apparent for the cases when the number of neighbors (n) for these water particles is $\le$ 4, whereas the presence of larger clusters (n$>$ 4) leads to a smaller tetrahedrality. 
The tetrahedrality (\textit{q}) autocorrelation function\cite{kumar_tetrahedral_2009}, $C_q(t)$ is given by

\begin{equation}
	C_q(t) = \frac{\left\langle q(t)q(0)\right\rangle-\left\langle q(0)\right\rangle^2}{\left\langle q(0)^2 \right\rangle - \left\langle q(0)\right\rangle ^2}\label{eqn:q_time}
\end{equation}

Here, the variance $\left\langle q(0)\right\rangle ^2$ is computed over all water particles over the entire trajectory, and the time-correlation of tetrahedrality $C_q(t)$ is measured with respect to several origins. It is noted that its decay for mW water model at 300 K occurs in $\sim$~10 ps: in very good agreement with the decay observed for atomistic water models\cite{kumar_tetrahedral_2009, garattoni_structure_2026}. Therefore, we expect the present coarse-grained water-glycan model to be capable of elucidating the role of glycans in disrupting the water tetrahedral rearrangements.

At 300 K, sufficient mobility exists due to liquid water even in the presence of CG8 chain. Hence, the $C_q(t)$ decay (teal) is almost equal to its decay seen for liquid water (blue).  At 180 K, a greater impact on the decay of $C_q(t)$ is observed for the CG8+water (purple) system than for pure water (magenta). Therefore, we can conclude that the CG8 chain inhibits the ability of water to rearrange towards a perfect tetrahedral ice-structure, as the decay $C_q(t)$ does not match its corresponding constant for pure water at 180 K. 

%Interestingly, we find that the excess kurtosis of the \textit{q} distribution for CG6, which is a measure of the fluctuation away from the peak (`tailedness') is larger than that of CG8. This ability to disrupt the \textit{q} distribution far from the peak may be understood as an indirect measure of the ability of cellulose to prevent ice-formation, since in the case of pure water, we expect the \textit{q} distribution to be a sharp peak at 1. 
%We hypothesize that this difference could be attributed to the ability of a given cellulose chain to have a greater curvature leading to a greater confinement effect as also discussed in section \ref{hydration}. 

% \begin{table}[h!]
% 	\caption{Excess kurtosis at 300 K and 180 K for water particles for different cellulose chain-lengths }\label{tab1}%
% 	\begin{tabular}{@{}llll@{}}
% 		\toprule
% 		Excess Kurtosis (K) & 4 units  & 6 units & 8 units\\
% 		\midrule
% 		300  &  1.20   & 1.27  & 1.28  \\
% 		180   & 7.17   & 7.80 & 6.48  \\
% 		$\Delta$   & 5.97   & 6.53  & 5.2\\%\footnotemark[2]  \\
% \botrule
% \end{tabular}
%\footnotetext{Source: This is an example of table footnote. This is an example of table footnote.}
%\footnotetext[1]{Example for a first table footnote. This is an example of table footnote.}
%\footnotetext[2]{Example for a second table footnote. This is an example of table footnote.}
%\end{table}

\noindent
\section{Conclusion}\label{sec3}

In this letter, we have demonstrated the disruption of tetrahedral linkage in water by cellulose-type glycans to be a key driver of their antifreeze activity, which is seen in the unfrozen water near the chains with ice-like character. This study combined with its predecessor \textit{ab initio} work, where we established that cellulose binds to ice planes using a tetrahedral coordination, together provide a fundamental picture of the tetrahedral structure of water/ice in their interaction with cellulose-derived materials.  As a result of these findings, we conclude that the ability of glycans to disrupt the tetrahedral structure of water is central to their antifreeze behavior, and serves as a design principle of derived synthetic antifreeze materials. 

\section{Materials and Methods}\label{sec14}
The MD simulations were carried out using the LAMMPS\cite{LAMMPS} package. To understand the key descriptors involved in antifreeze phenomenon, we utilize a simplified coarse-grained potential to simulate the freezing of glycan chains in water using classical molecular dynamics methods. We use the coarse-grained mW water\cite{molinero_water_2009} model that focuses on the tetrahedral structure of water and ice and has been successfully used to study freezing and ice formation\cite{sanchez-burgos_homogeneous_2022}. The glycan repeat unit was represented by a 12-6 Lennard-Jones potential represented by $\epsilon=0.2$ kcal/mol and $\sigma=4.5$~\AA. The bonded interactions within the chain were described by harmonic bonds and angles. Here, we also modeled the interaction potential between the chains and the mW water particles using a Lennard-Jones potential ($\epsilon$=1.0 kcal/mol and $\sigma=4.5$~\AA). With these parameters, we ensured that the cellulose-water interaction is hydrophilic. We used a timestep of 5 fs in all the simulations. 

Glycan chains of repeat varying lengths were initially equilibrated using the NPT ensemble with 27,648 water molecules at 300 K for 10 ns to obtain the initial starting points for our runs. This was followed by cooling the system at a rate of 0.1 K/ns to 180 K. For the analysis of tetrahedrality distribution of water for various cellulose-chains at 180 K, we subsequently conducted a production MD run using an NVT ensemble for 25 ns. The autocorrelation function was computed from the pure water and CG8+water systems, where an NPT ensemble was used to equilibrate the system for 20 ns, following which trajectories of 25 ps were obtained to compare their decays at 300 K and 180 K.

% {\color {magenta}
% \begin{equation}
% 	q_n = 1-\frac{9}{2n(n-1)} \sum_{j=1}^{n-1} \sum_{k=j+1}^{n}(\cos\psi_{jk}+\frac{1}{3} )^2 \label{eqn:tetra_all}
% \end{equation}
% }

\section*{Acknowledgements and Declaration of Conflicts and Other Disclosures}

{This project is supported by the United States Army Corps of Engineers, Engineer Research and Development Center’s Cold Regions Research and Engineering Laboratory (ERDC-CRREL, Contract No. W913E524C0009). Any opinions, findings, and conclusions or recommendations expressed in this material are those of the author(s) and do not necessarily reflect the views of the Broad Agency Announcement Program and ERDC-CRREL. The authors would like to thank Stony Brook Research Computing and Cyberinfrastructure and the Institute for Advanced Computational Science at Stony Brook University for access to the high-performance SeaWulf computing system, which was made possible by \$1.85M in grants from the National Science Foundation (awards 1531492 and 2215987) and matching funds from the Empire State Development’s Division of Science, Technology, and Innovation (NYSTAR) program (contract C210148).}

%\begin{itemize}
\backmatter
\bmhead{Conflict of interest/Competing interests}
The authors declare that there are no conflicts of interest. 

\bmhead{Data availability} 
All the data supporting the findings of this letter are available from the corresponding authors upon reasonable request.

\bmhead{Author contribution}
Credit: Aakash Kumar: conceptualization, methodology, software, validation, formal analysis, investigation, visualization, writing– original draft, writing– review and editing; Shoumik Saha: visualization, writing– review and editing; Dilip Gersappe: conceptualization, supervision, project administration, funding acquisition, resources, writing– review and editing.

\bibliography{sn-bibliography}% common bib file

\end{document}